\documentclass[aps,pre,twocolumn,showpacs,groupedaddress]{revtex4}

\usepackage{graphicx}
\usepackage{dcolumn}
\usepackage{bm}
\usepackage{epsfig}

\begin{document}


\title{Strongly nonlinear waves in capillary electrophoresis}

\author{Zhen Chen}
\author{Sandip Ghosal}%
 \email{s-ghosal@northwestern.edu}
\affiliation{%
Department of Mechanical Engineering, Northwestern University, Evanston, IL 60208
}%


\date{\today}

\begin{abstract}
In capillary electrophoresis, sample ions migrate along a micro-capillary filled with a background electrolyte 
under the influence of an applied electric field. If the  sample concentration is sufficiently high,
the electrical conductivity in the sample zone could differ significantly from the background. 
Under such conditions, the local migration velocity of sample ions becomes 
concentration dependent resulting in a nonlinear wave that exhibits shock like features.
If the nonlinearity is weak, the sample concentration profile, under certain simplifying assumptions, 
can be shown to obey 
Burgers' equation (S. Ghosal and Z. Chen {\it Bull. Math. Biol.} 2010 {\bf 72}(8),  pg. 2047)
which has an exact analytical solution for arbitrary initial condition.
In this paper, we use a numerical method to study the problem in the more general case 
where the sample concentration is not 
small in comparison to the concentration of background ions. In the case of low concentrations, 
the numerical results agree with the weakly nonlinear theory presented earlier, but at high concentrations, the wave 
evolves in a way that is qualitatively different. 
\end{abstract}
\maketitle

\section{Introduction} 
\label{sec:Intro}
In capillary electrophoresis (CE), separation of charged molecular species is accomplished by exploiting the 
differential migration of ions in a narrow channel (10--100 $\mu$m) in which a strong electric field ($\sim 100$ V/m) 
is applied in the axial direction~\citep{czebook1,czebook2}. 
The sample ions exist in solution in an electrolytic buffer which is referred to as the background electrolyte (BGE).
Separation is accompanied by the competing process of diffusive spreading in the axial direction which causes 
peak dispersion. Dispersion reduces resolution of the separation and may lower the peak concentration to below 
the detection threshold. It is therefore detrimental. Any effect that tends to increase axial spreading over the minimum 
imposed purely by molecular diffusion in the axial direction 
is referred to as ``anomalous dispersion''~\citep{ghosal_annrev06}. The transport problem of ions 
in the capillary is of considerable interest as it determines the amount of dispersion of the sample peak. 

In this paper we are concerned with an effect known as ``electromigration dispersion'' (EMD) that causes 
significant anomalous dispersion when the ratio of sample to background ion concentration becomes large enough.
For this reason it is also known as the ``sample overloading effect''. In CE, it is desirable to have the sample 
concentration at the inlet as high as possible (to ensure that even trace components are within detectable limits) 
and buffer conductivity as low as possible (to minimize Joule heat), so that the limitation imposed by EMD 
quickly becomes significant~\cite{jorgenson_lukacs_ac_81}. 

The physical mechanism of EMD may be explained roughly in the following way: when the concentration of sample ions is sufficiently high 
in comparison to that of the background electrolyte, the local electrical conductivity of the solution 
is altered in the region around the sample peak. However, charge conservation requires the electric current
to be the same at all points along the axis of the capillary. If diffusion currents due to concentration inhomogeneities are ignored 
for the moment, it follows, that the electric field must change axially. This is because Ohm's law, taken together with current conservation, 
implies that  the product of the conductivity and electric field 
must remain constant along the capillary. The axially varying electric field then alters the effective migration speed of the sample ions, which in turn alters 
its concentration distribution. Thus, in the continuum limit, the concentration of sample ions is described by a nonlinear transport 
equation. As expected, the CE signal exhibits features reminiscent of nonlinear waves familiar from other physical 
contexts~\cite{ghosal_chen10,whitam_book}. 

A one dimensional nonlinear hyperbolic equation for the sample ion concentration may be derived 
using simplifications that arise from assuming local electroneutrality and from neglecting 
the diffusivity of ions~\cite{mi_ev_ve_79a}. The restriction to zero ionic diffusivities was recently removed by Ghosal and 
Chen~\cite{ghosal_chen10}. They considered the minimal model  of a three ion system -- the sample ion, a 
co-ion and counter-ion. The diffusivities of the three ionic species were assumed equal, though not necessarily zero.
The sample ion concentration was then shown to obey a one dimensional nonlinear advection-diffusion equation which reduced to Burgers' 
equation if the sample concentration was not too high relative to that of the background ions.

In this paper, we focus on the minimal three ion system considered by Ghosal and Chen~\cite{ghosal_chen10} but we 
do not assume that the concentration of sample ions is small. Local electro-neutrality is however an excellent approximation 
in CE systems, since characteristic length scales are much larger than the Debye length which is on the order of 
nanometers. We therefore exploit it to reduce the numerical stiffness of the coupled ion transport equations. We identify a small number 
of parameters that primarily determine the system evolution and study the dynamics for a representative range of these parameters. 
We show that at low concentrations 
the peak evolves in accordance with the weakly nonlinear theory~\cite{ghosal_chen10}, but at high enough concentrations, 
the dynamics of peak evolution is qualitatively different as the system is dominated by the
nonlinearity. Surprisingly, in the strongly nonlinear regime, the peak breaks up into two zones marked by a critical concentration 
($\phi = \phi_c$) and separated by a diffusive boundary.  The high concentration zone ($\phi > \phi_c$) remains quasi-stationary 
whereas the low concentration zone propagates forward forming a ``surge front'' superficially resembling nonlinear wave phenomena 
familiar in the context of water waves, such as a river bore~\cite{stoker_formation_1948}.
 The critical concentration ($\phi_c$) can be predicted by a simple argument based on flux conservation.  At late times,
dispersion ensures that concentrations throughout the domain get smaller and the peak once again may be described by 
Burgers' equation. The complex nonlinear behavior is a consequence of the nonlinearity inherent in the Nerst-Planck 
equations of ion transport, just as the behavior of large amplitude water waves arise from the nonlinear nature of the Navier-Stoke's equations of hydrodynamics.

\section{Numerical Simulations}
We set up and solve numerically an idealized problem in which a sample peak migrates in a background electrolyte. 
 The channel walls are assumed charge neutral, so that electro-osmotic flow
is absent~\footnote{The effect of a wall zeta potential has recently been investigated~\cite{ghosal_chen_emd_eof}.}.
Further, local electro-neutrality is invoked which enables us to express the electric field in terms of the instantaneous 
concentration distributions rather than solve the Poisson's equation for the electric potential. This considerably simplifies the numerical 
work as the Poisson's equation is stiff on account of the smallness of the Debye length. Thus, the problem is reduced to solving a set 
of one dimensional coupled partial differential equations for the ion concentration fields.

\subsection{Model System} 
We will consider a three ion system consisting of sample ions, co-ions and counter-ions. Results will be expressed
in terms of dimensionless variables: all lengths are in units of a characteristic length $w_0$ determined by the initial peak width, 
time is in units of $w_0/v$, where $v$ is the migration velocity of an isolated sample ion in the applied field ($E^{\infty}$) and
the electric potential is in units of $E^{\infty} w_0$. All concentrations are in units of $c_n^{\infty}$,
where $c_n^{\infty}$ is the concentration of negative ions in the background electrolyte. In order to define a minimal problem 
with the fewest possible parameters, we
assume that the mobility ($u$) is the same for all the species, and therefore, so is the diffusivity ($D$), in accordance 
with the Einstein relation ($D_{i}/u_{i} = D/u  = k_{B} T$ where $k_{B}$ is Boltzmann's constant and $T$ is the absolute temperature). Note however, since the valence $z_i$ are different, the electrophoretic mobilities of the 
species $\mu_{i} = z_{i} e u$ are not identical. Then the only parameters in the problem are
 $Pe = vw_0/D$, which may be regarded as a ``P\'{e}clet number'' based on the electromigration 
 velocity $v$, and the two valence ratios 
$z_n/z,z_p/z$, where $z_p$, $z_n$ and $z$ are respectively the valence of cations, anions and sample.
We present results for two values of the P\'{e}clet number: $Pe=100$ and $200$ and fix the valence ratio
at $z:z_p:z_n=1:2:-1$. For other values of these parameters the results are qualitatively similar. The parameter 
of greatest interest is the degree of sample loading or the amplitude of the initial peak. 
The shape of the wave is insensitive to initial conditions, so for convenience we take the initial peak 
shape to have a  rectangular~\footnote{to reduce numerical errors, the corners of the rectangle were slightly ``rounded'' 
by using a tan hyperbolic function.} profile of height $c_{m}$ and width $2 w_0$ centered at $x=10w_{0}$. This is also the most common initial shape encountered in practice where the sample is introduced by electrokinetic injection. The degree of sample loading 
is conveniently characterized \cite{ghosal_chen10} in terms of the quantity 
\begin{equation} 
\Gamma = \int_{-\infty}^{+\infty} \phi(x,t) \; dx =  \int_{-\infty}^{+\infty} \frac{c_n}{c_n^{\infty}} \; dx,
\end{equation} 
which has units of length. The length scale $\Gamma$ may be used to define a second P\'{e}clet number $P = v \Gamma /D$ 
which may be treated as a dimensionless measure of sample loading. 
A series of simulations are conducted with peak heights in the range $\phi_{m} = c_{m}/c^{\infty}_n = 0.01$  (low sample loading) to $0.8$ (high sample loading). The initial 
co-ion concentration $c_p$ is assumed constant throughout the domain. Then the counter-ion concentration is determined 
by the local electro-neutrality constraint, Eq.~(\ref{eq:LEN}). 
An infinite domain is approximated by a finite computational box of length much greater than $w_{0}$. The values of the 
concentrations are held fixed at the domain boundaries and $\partial \phi_e / \partial x$ is set to the constant value 
$-E^{\infty}$.
The domain is chosen to be sufficiently large that the perturbations of the concentrations and fields are always negligible near the domain boundaries.

\subsection{Numerical Method} 

We will solve the governing equations for ion transport  in solution, which  are
\begin{equation}
	\frac{\partial c_{i}}{\partial t} + \frac{\partial}{\partial x}\left[ -\mu_{i} c_{i} \frac{\partial \phi_e}{\partial x} - D_{i} \frac{\partial c_{i}}{\partial x}  \right]= 0
	\label{eq:transport}
\end{equation}
where $c_{i}$ is the concentration of species $i$ ($i=1,2,\ldots,N$) with electrophoretic mobility $\mu_{i}$ and diffusivity $D_{i}$. Electro-osmotic 
flow is neglected so that the problem is one dimensional and may be described using the co-ordinate 
$x$ along the capillary and time since injection, $t$. On account of the requirement of local electroneutrality~\cite{rubinstein_book}
 \begin{equation} 
\sum_{i=1}^{N} z_{i}  c_{i} = 0
\label{eq:LEN}
\end{equation} 
($z_i$ is the valence of the $i$th species).
The electric potential 
 $\phi_e$ may be found from the equation of current conservation: 
\begin{equation}
	\frac{\partial}{\partial x}\left[ -\sum^N_{i = 1} z_i  \mu_{i} c_{i} \frac{\partial \phi_e}{\partial x} - \sum^N_{i = 1} z_iD_{i} \frac{\partial c_{i}}{\partial x}  \right]= 0.
	\label{eq:current}
\end{equation}
Eq.~(\ref{eq:current}) may be readily integrated to yield the local electric field, $E = - \partial_{x} \phi_{e}$:
\begin{equation} 
E(x,t) = \frac{E^{\infty} \sum_{i} z_i  \mu_{i} c_{i}^{\infty} + 
\sum_{i} z_i  D_{i} \partial_{x} c_{i}}{\sum_{i} z_i  \mu_{i} c_{i}},
\label{eq:E(x)}
\end{equation}
where the superscript $\infty$ indicates the value of the respective variable far away from the peak
and the summation is over all species.
 
A finite volume method is used to discretize equations (\ref{eq:transport}) and (\ref{eq:current})  in space using 
an adaptive grid refinement algorithm that is enabled by applying the Matlab library ``MatMOL''~\cite{vande_wouwer_simulation_2004}.
The spatially discretized system of equations is then integrated in time using the Matlab solver ``ode45''~\cite{shampine_matlab_1997}
which is based on an explicit Runge-Kutta (4,5) formula. Equations (\ref{eq:transport}) and (\ref{eq:current}) automatically ensure 
that the electro-neutrality condition, Eq.~(\ref{eq:LEN}), is satisfied and this is verified at each time step. 

\subsection{Results} 
\begin{figure}[t]
    \begin{center}
  \includegraphics[width = 0.5\textwidth,angle=0]{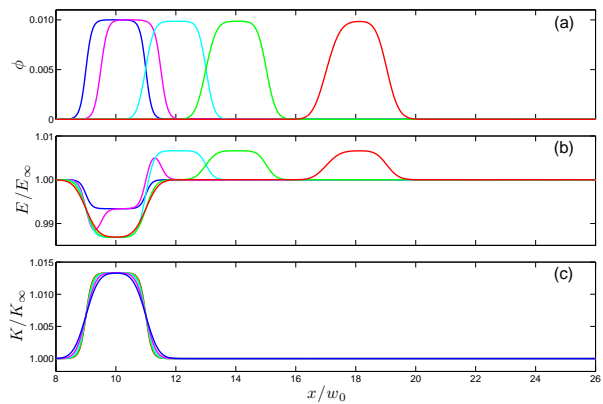}  
    \caption{Time evolution of the profiles of the normalized sample ion concentration ($\phi$), electric field ($E/E_{\infty}$)
and Kohlrauch function ($K/K_{\infty}$) in the case of weak sample loading ($\phi \ll \phi_{c}$). The Kohlrausch function spreads only by diffusion so that the sample peak rapidly migrates into the zone where $K = K_{\infty}$
(movie online).}
    \end{center}
\end{figure}
\begin{figure}[t]
  \begin{center}
        \includegraphics[width = 0.5\textwidth, angle = 0] {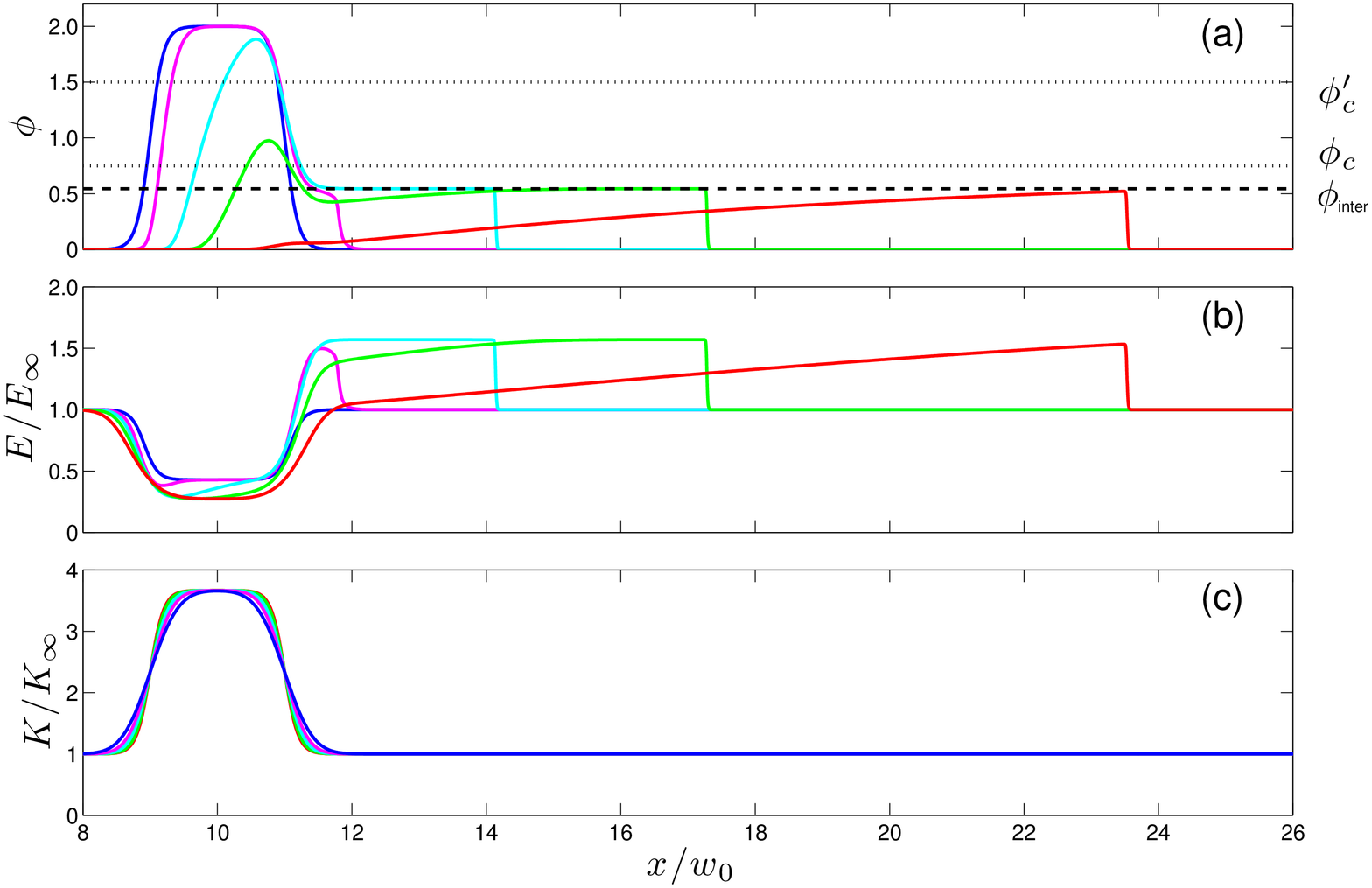}   
    \caption{The same as in Fig.~1 except here the amplitude of $\phi$ exceeds  $\phi_c$. Here the part 
    of the peak above the
value $\phi = \phi^{inter}$ appears to be effectively immobilized. 
The middle panel shows that the stagnant zone is due to a sharp reduction in the electric field 
caused by the very high electrical conductivity in this zone. The assumption $K=K_{\infty}$ is clearly invalid as a part of the peak 
remains trapped in the injection zone (movie online).}
  \end{center}
\end{figure}

Figure~1(a) shows the profiles of the normalized sample concentration $\phi(x,t) = c_n/c_n^{\infty}$ at fixed times 
$vt/w_{0}=0,0.5,2.0,4.0$ and $8.0$ for the case of low sample loading. Fig.~1(b) and (c) show respectively 
 the profiles of the corresponding electric field 
$E(x,t)$ and the Kohlrausch regulating function $K(x,t) = (c_p + c_n + c)/u$. 
The Kohlrausch regulating function is a useful quantity for describing electrokinetic transport.  If all ionic 
species have the same diffusivity, $K(x,t)$ evolves as a passive scalar \cite{ghosal_chen10}. If ionic diffusivities are treated as zero, then $K(x,t)$ 
is a conserved quantity~\cite{kohlrausch}. 
It is seen that $K(x,t)$ remains localized near the 
injection zone and spreads only slowly by molecular diffusion.
The sample peak on the other hand moves to the right and after a short 
time, the sample peak essentially lies in a zone where $K=K_{\infty}$. This illustrates the behavior postulated earlier 
that makes possible a simplified description in terms of the one dimensional nonlinear equation~\cite{ghosal_chen10}:
\begin{equation} 
\frac{\partial \phi}{\partial t} + \frac{\partial}{\partial x} \left( \frac{v \phi}{1 - \alpha \phi}  \right)
= D \frac{\partial^{2} \phi }{\partial x^{2}}.
\label{1Dnlwave}
\end{equation} 
If $\phi$ is small, Eq.(\ref{1Dnlwave})
reduces to the Burgers' equation on Taylor expansion of $(1 - \alpha \phi)^{-1}$.
In the vicinity of the sample peak, the electric field 
 is functionally related to the normalized sample concentration; $E = E^{\infty}/(1 - \alpha \phi)$. Here
 $\alpha$ is the ``velocity slope parameter'' introduced in \cite{ghosal_chen10}.

It may be shown~\cite{ghosal_chen10} that the requirement of positivity of co- and counter-ion concentrations 
implies that only sample profiles satisfying the condition $\phi < \phi_c$, where $\phi_c$ is a positive 
number, may be described by the theory. We will call such profiles ``realizable''.
The critical concentration, $\phi_c$,  is given by $\phi_{c} = (z_p - z_n)/(z_p - z)$ when 
$z < 0$ and $\phi_{c} = - [ z_n (z_p - z_n) / [ z_p (z - z_n) ]$ when $z > 0$. When the parameter $\alpha >0$, it may be shown with some simple algebra that $\phi_c < \phi_c^{\prime} \equiv \alpha^{-1}$ (see Appendix),  so that the 
singularity implicit in Eq.(\ref{1Dnlwave}) when 
$\phi = \phi_c^{\prime}$ is never reached for realizable solutions. Fig.~2 shows the behavior of the system for initial conditions that are not realizable. 
In this situation, a stationary ``barrier'' develops at a fixed spatial location corresponding to a certain value 
$\phi = \phi^{inter} < \phi_{c}$. 
The sample ions move more or 
less freely on crossing the barrier but are effectively immobilized on the left of the barrier. 
This is due to the greatly reduced strength of the electric field 
in the injection zone where the electrical conductivity is high.
This is clearly seen in Fig.~2(b) which shows a sharp reduction in the electric 
field in the injection zone. Only sample ions near the edges 
of the zone are able to ``leak out'' and are carried to the right as an advancing wave. Since part of the sample profile 
remains quasi stationary, the assumption of the constancy of the Kohlrausch function, $K = K_{\infty}$ can no longer be made for non-realizable concentrations. Thus, Eq.(\ref{1Dnlwave}), which would have led to 
unphysical negative concentrations for such non-realizable profiles, is not applicable until after a sufficient 
time has evolved so that $\phi$ is reduced to a value below $\phi^{inter}$ throughout the domain. 

\begin{figure}[t]
    \begin{center}
    \includegraphics[width = 0.5\textwidth,angle=0]{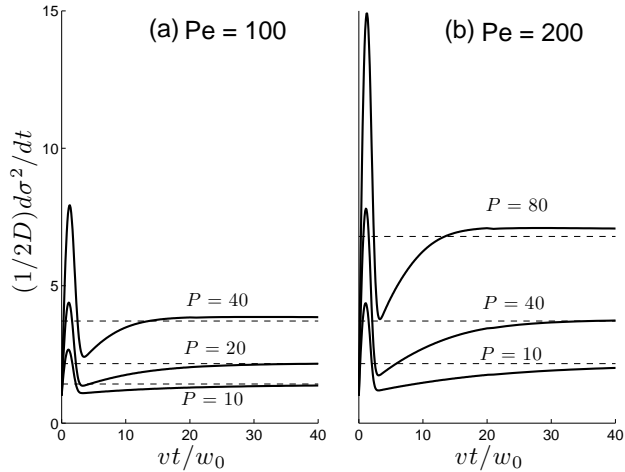}       
    \caption{The normalized rate of change of variance as a function of dimensionless time 
    $vt/w_{0}$ for (a) $Pe=100$  and (b) $Pe=200$  for three different values of sample loading ($P$). 
At long times, the peak is seen to 
    spread with an effective diffusivity 
    $D_{\mbox{eff}} = \lim_{t \rightarrow \infty} (2D)^{-1} (d \sigma^{2} / dt)$ 
    given by Eq.~(37) of \cite{ghosal_chen10} and indicated here by the horizontal dashed line.
    }
    \end{center}
\end{figure}
\begin{figure}[t]
    \begin{center}
    \includegraphics[width = 0.5\textwidth,angle=0]{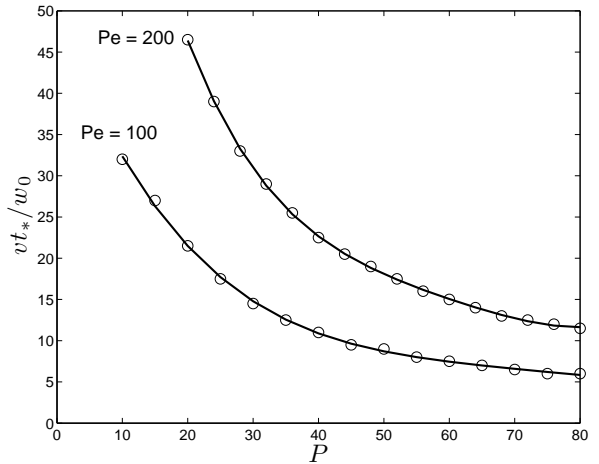}
    \caption{The normalized convergence time $vt_{*}/w_{0}$ where $t_{*}$ is the time taken for the effective variance $(2D)^{-1} (d \sigma^{2} / dt)$ to reach $95$ percent of its final asymptotic value of $D_{\mbox{eff}}$ as a function 
    of the sample loading ($P$) for different values of the P\'{e}clet number ($\text{Pe}$).}
    \end{center}
\end{figure}
Fig.~3 shows the variation in time of the quantity $(2D)^{-1} d \sigma^{2} / dt$ for a series 
of different values of the diffusivity and sample loading  characterized by the pair of P\'{e}clet numbers 
$(\text{Pe},P)$.
If the profile spread purely by molecular diffusivity, this quantity should approach one asymptotically. However, 
it is seen that the long time asymptotic value is not one but rather $D_{\mbox{eff}}$ which depends solely 
on $P$. The dashed line shows the theoretical value of $D_{\mbox{eff}}$ predicted by the weakly nonlinear 
theory based on solutions of the Burgers' equation~\cite{ghosal_chen10}. Thus, once the system has evolved 
long enough, and dispersion has caused the amplitude to drop sufficiently, Burgers' equation provides a 
valid  description of the peak evolution. However, a real separation happens in a finite capillary 
and the long time limit may not necessarily apply. A quantity of interest is the timescale characterized by 
$t_{*}$: the time needed for the quantity $(2D)^{-1} d \sigma^{2} / dt$ to relax to $0.95$ of its asymptotic 
value $D_{\mbox{eff}}$. If the separation is conducted in a capillary of length $L$, the question of interest 
is whether $t_{*}$ is small or large compared to the total separation time $T = L/v$. In Fig.~4
we show the normalized time  $v t_{*} /w_{0}$ from a series of simulations with different values of $(Pe,P)$.
Clearly, $v t_{*}/w_{0}$ is a monotonically decreasing function of $P$. This can be anticipated from the 
theory of nonlinear waves~\cite{whitam_book}: the higher the amplitude, the quicker a shock or shock like 
structure is formed. 
In contrast to the effective diffusivity shown in Fig.~3 which depends on $P$ but not on $Pe$,
the time to reach the asymptotic state does depend on $Pe$. In fact, as Fig.~4 shows, the 
curve $v t_{*} /w_{0}$ as a function of $P$ is shifted upwards as $Pe$ is increased. Indeed, larger 
$Pe$ corresponds to lower diffusivity and therefore a longer time for the peak to spread and 
its amplitude to fall sufficiently for the weakly nonlinear description to be valid. Typical values of the 
physical parameters in a microchip based system may be $w_{0} \sim 100 \mu$m, $L \sim 5$ cm,
so that $v T/w_{0} \sim 500$. Thus, Fig.~4 suggests that the Burgers' solution does 
describe the peak dynamics for most of the separation time except for possibly a relatively short 
initial transient.

\subsection{Analysis} 
\begin{figure}[t]
\begin{center}
 \includegraphics[width =0.5\textwidth, angle =0]{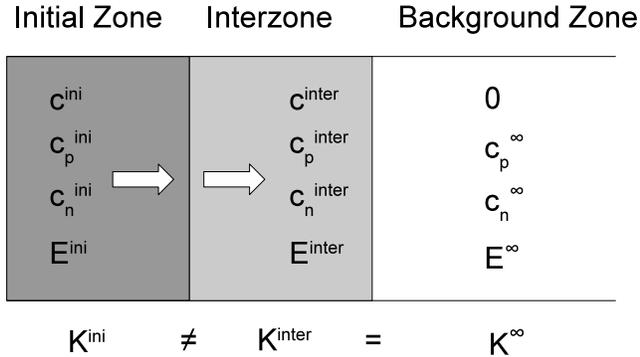}
 \caption{Schematic diagram describing approximately the initial phase of peak evolution when $\phi > \phi_{c}$. The 
domain can be divided into an ``Initial Zone'' an ``Interzone'' and a ``Background Zone''. All the dependent variables are approximately 
constant within each zone but may undergo jumps across zone boundaries.}
 \end{center}
\end{figure}
An approximate theoretical determination of the concentration  $\phi^{inter}$ may be provided using the conservation 
equations. The method of doing this is in fact entirely analogous to the ``Moving Boundary Equations'' (MBE)~\cite{dole_theory_1945} 
for describing advancing fronts (e.g. in isotachophoresis), except, in this case, the front happens to be quasi stationary. The conceptual framework is 
illustrated in Fig.~3. The domain is decomposed into three parts: the ``Initial Zone'' where the sample is injected, the ``Background Zone''
ahead of the advancing wave where all concentrations equal their initial values and an ``Interzone'' between them. All variables 
are assumed constant within each zone but undergo a discontinuous change across zone boundaries. The values of 
the variables in each zone are indicated in Fig.~3. The boundary between the Initial Zone and the Interzone is stationary 
whereas the boundary between the Interzone and the Background Zone moves to the right. The arrows indicate fluxes 
of ions across the stationary zone boundary. 
Conservation of these ionic fluxes require 
\begin{eqnarray}
	E^{ini}\phi^{ini} &=& E^{inter}\phi^{inter}\\
	E^{ini}\phi^{ini}_n &=& E^{inter}\phi^{inter}_n
\end{eqnarray}
where $E$ represents the electric field and $\phi$ represents the concentration (normalized by $c_{n}^{\infty}$).
The superscript (``ini'' for the Initial Zone, ``inter'' for the Interzone and ``$\infty$'' for the Background Zone) indicates the zone in which the variable is evaluated and the 
subscript ($p$ for cation, $n$ for anion and no subscript for the sample) identifies the species.
Therefore,
\begin{equation}
	\phi^{inter}_n = ( \phi^{ini}_n / \phi^{ini} ) \phi^{inter}
	\label{eq:phi_n}
\end{equation}
For the inter zone, 
\begin{eqnarray}
	K^{inter} &=& c_{n}^\infty  \left( \phi^{inter} + \phi^{inter}_p + \phi^{inter}_n \right) / u \nonumber\\
	&=& K_{\infty} =  c^{\infty}_n \left( \phi^{\infty}_p +  1 \right) / u  \nonumber\\
	&=& c^{\infty}_n  \left( 1  - z_n / z_p \right) /u.
	 \label{eq:Kinter}
\end{eqnarray}
The electro-neutrality condition (valid in all zones) is:
\begin{equation} 
z_p \phi_p + z_n \phi_n + z \phi = 0. 
\end{equation} 
By combining Eq.~(\ref{eq:phi_n}) and (\ref{eq:Kinter}) and using the electro-neutrality condition 
we get  an equation for determining $\phi^{inter}$
\begin{equation}
	(1 - z/z_p) \phi^{inter} + r (1 - z_n/z_p) \phi^{inter}  = 1 - z_n/z_p,
\end{equation}
where the ratio $\phi^{ini}_n / \phi^{ini} = r$ is a constant determined by the ionic composition of the injected zone. 
Solving the above linear equation for $\phi^{inter}$ we have 
\begin{equation}
	\phi^{inter} = \left[ r + (z_p - z)/(z_p - z_n) \right]^{-1}.
	\label{eq: phi inter}
\end{equation}
In our numerical experiment the cation concentration was chosen to be uniform, so that 
$\phi^{ini}_p = \phi^{\infty}_p = -z_n/z_p$. Since $z:z_p:z_n=1:2:-1$, $r = 1/\phi^{ini} -z/z_n = 1.5$, and,  $\phi^{inter} = 0.55$. 
This value is indicated by the dashed line in Fig.~2(a).  Clearly, it correctly describes the concentration of sample in the Interzone. 

Thus, the theoretical description developed in \cite{ghosal_chen10} 
may be used in the Interzone 
($\phi < \phi^{inter}$) but not in the Initial Zone. In order that all ion concentrations be non-negative in the Interzone we
 must have $\phi^{inter} < \phi_{c} < \phi_{c}^{\prime}$. 
This inequality is indeed true as can be shown by some simple algebra (see Appendix).
 
\section{Conclusions}
The development of nonlinear waves in capillary electrophoresis in the limit of low as well as high 
 concentration of sample ions was studied by numerical integration of the governing equations. An idealized 
 minimal model was considered
consisting  of a three ion (sample, co-ion and counter-ion) system of strong 
electrolytes~\footnote{the situation of a weak electrolytic buffer was recently investigated by the authors~\cite{EMD1}.}. 
This study complements an earlier paper by the authors.  There it was shown
that, in the weakly nonlinear limit, the evolution of the sample concentration
may be reduced to Burgers' equation, which admits an exact analytical solution.

Numerical simulation revealed that the evolution of the peak proceeds in a way that is qualitatively 
different when the sample concentration is high.
 As a consequence of the sharp reduction of the electric field in the region of sample 
injection, the ion migration velocity in this zone is very small.  Ahead of  this zone the ions 
form a surge front with a step-like profile propagating to the right.
This state of affairs continues until the dimensionless ion concentration ($\phi$)
in the injection zone drops sufficiently so that $\phi < \phi^{inter}$. The subsequent dynamics 
then proceeds in accordance with the weakly nonlinear theory~\cite{ghosal_chen10}.  The value of $\phi^{inter}$
may be approximately calculated by using a simple model based on conservation of ionic fluxes. 

This qualitative change in the dynamics of peak evolution explains the breakdown of the 
weakly nonlinear theory  
when the concentration $\phi$ exceeds the critical value $\phi_c$. When $\phi$ exceeds a certain 
value $\phi^{inter} < \phi_{c}$
part of the propagating wave is effectively immobilized in the injection zone. It is then no longer correct to 
assume~\cite{ghosal_chen10} that the sample pulse would quickly move out to a region where the 
Kohlrausch function is constant. 

The model studied here is clearly oversimplified. In particular, real electrophoresis buffers contain many more than three ions
including one or more weak acids or bases to maintain a stable pH. Further, complex effects due to inhomogeneities in the 
electroosmotic flow may be relevant~\cite{chen_effect_2008}. In this paper we ignore these complexities and attempt 
to produce a detailed understanding of a ``minimal'' model problem. One may 
question whether the strongly nonlinear regime considered here is of relevance to actual laboratory practice. The answer
depends on the numerical values of the critical concentrations $\phi^{inter} < \phi_{c} < \phi_{c}^{\prime}$. 
If the sample and carrier ions have similar valences then all of these critical concentrations are of order unity. Thus, to 
exceed these critical values the sample ions in the injected plug will need to be present at concentrations approaching 
that of the carrier electrolyte. Such high concentrations are rarely employed in laboratory practice. However, if the sample 
is a macro-ion the critical values may actually be quite small. For example, at pH 2.0 Bovine serum albumin has a 
valence, $z \sim 55$~\cite{ford_measurement_1982}. Then, in a univalent 
carrier electrolyte we have $\phi_{c} \sim 0.04$, so that the strongly nonlinear regime studied here may be easily reached.\\
\begin{acknowledgments} 
This work was supported by the National Institute of Health  under grant R01EB007596.
\end{acknowledgments}

\bibliographystyle{osa}

\appendix*
 \setcounter{equation}{0}  
 \section*{Appendix: Proof of the inequality $\phi^{inter} < \phi_{c} < \phi_{c}^{\prime}$}

The critical concentration $\phi_c$ is defined as~\cite{ghosal_chen10}
\begin{equation}
	\phi_c = \left\{
	\begin{array}{ll}
	\frac{z_p - z_n}{z_p - z} & \mbox{if $z<0$}\\
	- \frac{z_n}{z_p} \frac{z_p-z_n}{z - z_n} & \mbox{if $z>0$}
	\end{array}
		\right.
\end{equation}
whereas 
\begin{equation} 
\phi_{c}^{\prime} = \frac{1}{\alpha} = \frac{z_n (z_p - z_n )}{(z-z_n)(z - z_p)}.
\end{equation}
We need to show that $\phi_c < \phi_c^{\prime}$ when $\alpha > 0$, that is, when $z_p > z > z_n$.
To do this, evaluate the ratio $\phi_{c}/\phi_{c}^{\prime}$ when $z_p > z > z_n$:
\begin{equation}
	\frac{\phi_c}{\phi_{c}^{\prime}} = \left\{
	\begin{array}{ll}
	\frac{z_p - z_n}{z_p - z}  \frac{(z-z_n)(z - z_p)}{z_n (z_p - z_n )} = \frac{-z_n + z}{-z_n} < 1& \mbox{if $z<0$}\\
	- \frac{z_n}{z_p} \cdot \frac{z_p-z_n}{z - z_n} \frac{(z-z_n)(z - z_p)}{z_n (z_p - z_n )} = \frac{z_p - z}{z_p} < 1& \mbox{if $z>0$}
	\end{array}
		\right.
\end{equation}
which completes the proof. 

To prove the remaining inequality, $\phi^{inter} < \phi_{c}$, we first show that $r > -z /z_n$ when $z<0$. To do this, we use the electro-neutrality 
condition to express $\phi_{p}^{ini}$ in terms of the other variables 
\begin{equation} 
\phi_{p}^{ini} = - \frac{z}{z_p} \phi^{ini} - \frac{z_n}{z_p} \phi_{n}^{ini}  = \frac{\phi^{ini}}{z_p} (- z - r z_n ).
\end{equation} \\[1ex]
\noindent Now we must have $\phi_{p}^{ini} > 0$. This is always true if $z<0$, but if $z>0$ then we require that $r > - z/z_n$. 

First suppose that $z <0$. Then 
\begin{eqnarray} 
\phi^{inter} &=& \frac{1}{ r + (z_p - z)/(z_p - z_n) } \nonumber \\
&<& \frac{1}{ (z_p - z)/(z_p - z_n) } \nonumber \\
&=& \frac{z_p - z_n}{z_p - z} = \phi_{c}
\end{eqnarray}

Now suppose that $z >0$. Then 
\begin{eqnarray} 
\phi^{inter} &=& \frac{1}{ r + (z_p - z)/(z_p - z_n) } \nonumber \\
&<& \frac{1}{ -(z/z_n) + (z_p - z)/(z_p - z_n) } \nonumber \\
&=& - \frac{z_n}{z_p} \frac{z_p - z_n}{z - z_n} = \phi_{c}
\end{eqnarray}
Thus, in all cases, $\phi^{inter} < \phi_{c}$ which completes the proof.

\end{document}